\documentclass[apj,iop]{emulateapj}
\usepackage{apjfonts, graphicx, comment}
\usepackage[normalem]{ulem}
\usepackage{color, soul}

\begin{document}

\title{A \textit{NuSTAR} observation of the low mass X-ray binary GX~349+2 throughout the Z-track}
\shortauthors{Coughenour et al.}
\shorttitle{GX~349+2 throughout its Z-track}

\author{Benjamin~M.~Coughenour\altaffilmark{1}}
\author{Edward~M.~Cackett\altaffilmark{1}}
\author{Jon~M.~Miller\altaffilmark{2}}
\author{Renee~M.~Ludlam\altaffilmark{2}}

\email{coughenour@wayne.edu}

\affil{\altaffilmark{1}Department of Physics \& Astronomy, Wayne State University, 666 W. Hancock St, Detroit, MI 48201, USA}
\affil{\altaffilmark{2}Department of Astronomy, University of Michigan, 1085 South University Ave, Ann Arbor, MI 48109-1107, USA}

\begin{abstract} 

Although the most luminous class of neutron star low mass X-ray binaries, known as Z sources, have been well studied, their behavior is not fully understood. In particular, what causes these sources to trace out the characteristic Z-shaped pattern on color-color or hardness-intensity diagrams is not well known. By studying the physical properties of the different spectral states of these sources, we may better understand such variability. With that goal in mind, we present a recent \textit{NuSTAR} observation of the Z source GX~349+2, which spans approximately 2 days, and covers all its spectral states. By creating a hardness-intensity diagram we were able to extract five spectra and trace the change in spectral parameters throughout the Z-track. GX~349+2 shows a strong, broad Fe K$\alpha$ line in all states, regardless of the continuum model used. Through modeling of the reflection spectrum and Fe K$\alpha$ line we find that in most states the inner disk radius is consistent with remaining unchanged at an average radius of 17.5 $R_g$ or 36.4 km for a canonical 1.4 $M_\odot$ neutron star. During the brightest flaring branch, however, the inner disk radius from reflection is not well constrained.

\end{abstract}

\section{Introduction}

Low mass X-ray binaries (LMXBs) are systems that consist of a compact object as well as a low mass ($\lesssim$ 1 $M_\odot$) companion star. Although LMXBs have been well studied, there is still a lot of complexity about the variability in these systems that is not well understood. Neutron star (NS) LMXBs can vary over orders of magnitude in luminosity, and are classified by their output as well as their variability. The two main groups are the less luminous Atoll sources, which output roughly 1 to 10 percent of their Eddington luminosity, and the much more luminous Z sources, which have high enough accretion rates to radiate an appreciable fraction of their Eddington luminosity --- from 50 percent to 100 percent or more \citep{vanderKlisbook2005}. The two classes are primarily differentiated by the distinct shapes they form when plotted in color-color or hardness-intensity diagrams \citep[see][]{HasingerVDK1989}.

Z sources are so named because of the Z-shaped track they trace out on the hardness-intensity diagram (HID) over the course of one or several days \citep{HasingerVDK1989}. The three distinct branches are named the horizontal, normal, and flaring branches (HB, NB, and FB, respectively). These sources can be further separated into two sub-groups: the Scorpius~X-1 like sources, which include alongside Sco~X-1 the sources GX~349+2 and GX~17+2 \citep{Church_etal2012}, and the Cygnus~X-2 like sources, which include Cyg~X-2, GX~5-1 and GX~340+0. Sco-like sources exhibit little or no HB and strong flaring, while the Cyg-like sources show a strong HB but have weak flaring \citep{Church&Church2012}. One transient source, XTE~J1701-462, has been observed to transition from a Cyg-like Z source to an Atoll source \citep{Lin_etal2009, Homan_vanDK_etal2010}. In fact, \citet{Homan_vanDK_etal2010} suggested that the source transitioned intermediately into a Sco-like Z source during their observation, before becoming an Atoll source. However the Sco-like section of their data did not display NS temperatures greater than 2 keV, which is characteristic of Sco-like sources \citep{Church_etal2012}. In any case, the transient source XTE~J1701-462 demonstrates that there is no inherent difference in these two main types of NS LMXB sources other than environmental factors such as the rate of accretion \citep{Lin_etal2009, Homan_vanDK_etal2010}.

\begin{figure*}[hbt!]
\centering
\includegraphics[trim=3.0cm 2.2cm 3.5cm 3.0cm, clip=true, width=15cm]{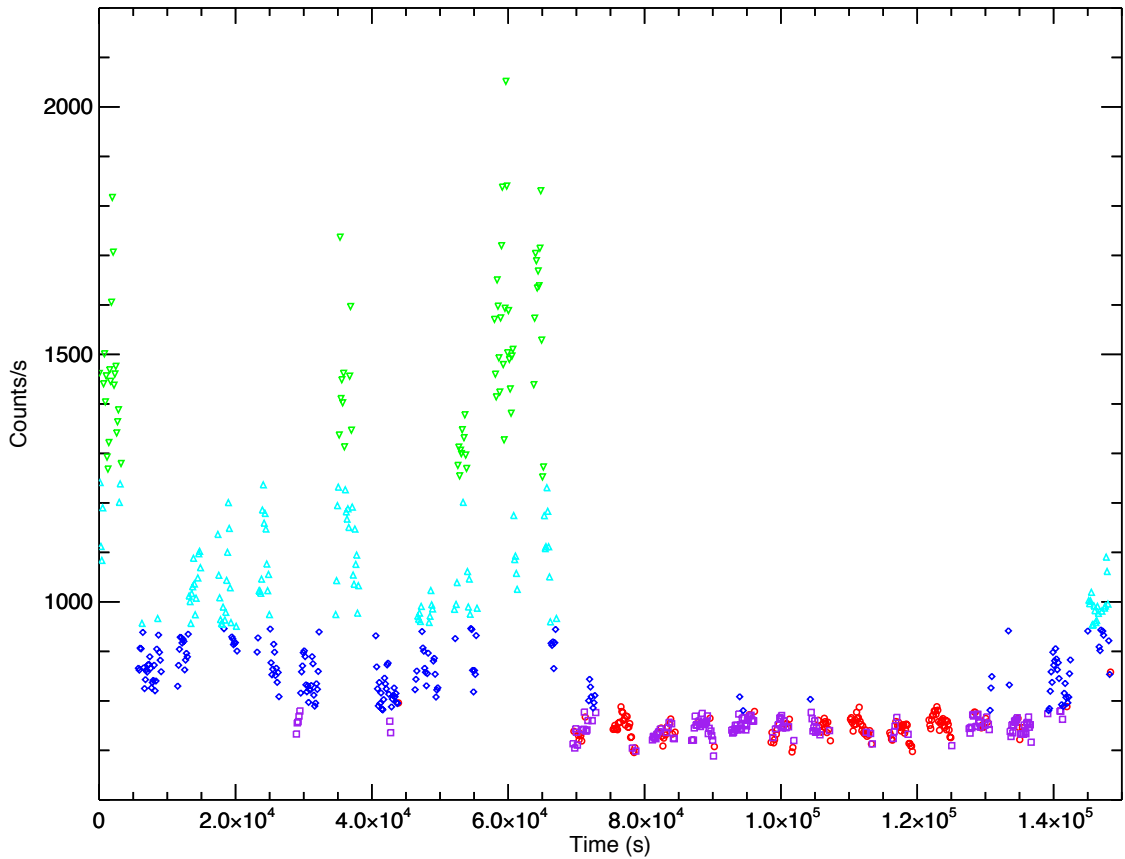}
\caption{Lightcurve of GX~349+2. The total 3 -- 50 keV light curve over the full 150 ks spanning the 80 ks observation is shown. Lightcurves for FPMA and FPMB were added together. Gaps in the lightcurve due to Earth occultations are clearly visible. The spectral states are shown as different colored shapes, with the normal branch in red (circles), the vertex in purple (squares), flaring branch 1 in blue (diamonds), flaring branch 2 in cyan (upward triangles), and flaring branch 3 in green (downward triangles). The flaring of the source here is clear. Error bars for these points are not shown, as they are smaller than the symbols used.}
\label{LC}
\end{figure*}

Although these Z-track LMXBs have been well studied, the physical properties that determine a source's spectral state, and which branch it occupies on its color-color or hardness-intentsity diagram, are still being understood. Furthermore, it is not well understood what actually drives a source to change its state and move about its Z-track. The classical argument is that changes in the mass accretion rate, $\dot{M}$, are responsible for a source changing its spectral state or moving along its branches in the HID, as proposed by \citet{Priedhorsky_etal1986}. Support for an increase of $\dot{M}$ from the HB to the NB, and then along the FB, came from a multi-wavelength study of Cyg X-2 by \citet{Hasinger_etal1990}. In the past decade, however, it has been shown that the variability of Z sources is not so simple as this \citep{Lin_etal2009, Church&Church2012}. For example, there is strong evidence to suggest that the mass accretion rate along the NB actually increases from the soft apex where the NB meets the FB, towards the hard apex where it joins the HB, since most of the increase in total luminosity towards the hard apex is not due to the properties of the NS. It may also be that the similar branches of Sco-like and Cyg-like sources, particularly flaring, are actually caused by different physical conditions \citep{Church_etal2012}. The obvious complexity of these systems motivates a better understanding of the physical differences between a Z source's different spectral states. 

\citet{Church&Church2012} identified that flaring in Cyg-like sources is due to unstable nuclear burning on the neutron star surface, citing an increase in blackbody temperature associated with the NS from spectral modeling. For Sco-like sources, however, flaring is much more prevalent. Furthermore, after finding much hotter blackbody temperatures in Sco-like sources at the vertex between the NB and FB, \citet{Church&Church2012} proposed that an increase in $\dot{M}$, along with unstable nuclear burning, may be responsible for the frequent flaring in Sco-like sources \citep{[see also] Church_etal2012}. Indeed, measured mass accretion rates for Z sources at the vertex between the NB and FB agree with the critical value of accretion flow defining the boundary between stable and unstable nuclear burning \citep{Church&Church2012, Church_etal2012}, as calculated by \citet{Fujimoto_etal1981} and \citet{Bildsten1998}. The link between unstable nuclear burning and flaring in Z sources is therefore strong.

An additional spectral probe is the broad Fe K line, which has recently been used to study several NS LMXB systems. The use of the Fe K$\alpha$ line as a tool for mapping the geometry of accreting systems first occurred in AGN and then in black hole (BH) binary systems \citep[see][for a review]{Miller2007}. More recently, broad Fe emission lines have been robustly detected and used as a means of understanding neutron star LMXBs as well \citep[e.g.][]{Asai_etal2000, diSalvo_etal2005,  diSalvo2009, Bhattacharyya2007, Cackett2008, Cackett2010, DAi2009, Chiang2016a}. Studying this emission line provides insight into accretion in the innermost region of these systems and may even be used to set an upper limit on NS radii \citep{Cackett2008, diSalvo2015, Chiang2016b, Ludlam2017a, Ludlam2017b}. Because the emission of the line is thought to originate in the inner accretion disk as the disk is exposed to hard X-ray photons from either a boundary layer or an extended corona, it is possible to infer details about the inner accretion disk as well as the geometry of the system by measuring the profile of the line. In this scenario, the line's profile is distinctly shaped by relativistic effects and the strong gravity of the neutron star \citep{Fabian1989}. Another signature feature from relativistic reflection would be the Compton backscattering hump at higher energies, though this is often difficult to detect. Free from the effects of photon pile-up, which can distort the profile of the Fe K emission lines, and with its broad energy range (3 -- 79~keV), the X-ray satellite \textit{NuSTAR} \citep{HarrisonNuSTAR2013} can be a powerful tool in ideal conditions for measuring the inner accretion disk radius and constraining neutron star radii \citep[see, e.g.,][]{Miller_etal2013, Degenaar_etal2015, King_etal2016, Ludlam2016, Ludlam_etal2017c, Sleator2016}.

The earliest works on LMXBs showed that Comptonization of thermal emission --- the upscattering of lower temperature photons by a hot plasma --- is required to account for the spectral shape of these systems, along with the thermal flux from either the neutron star or accretion disk, and that this Comptonization component often dominates the spectra, especially in the case of Atoll sources \citep{White_etal1988}. This is still an important factor in the spectra of Z sources as well, however its share of the total flux decreases in higher flux states \citep{Mitsuda_etal1989}. In any case, while a disk component is necessary for reflection modeling of a broad Fe K$\alpha$ line, the use of models that include Comptonization has lead to great successes in revealing changes in the continuum spectra of these Z sources as they alter their spectral shape, as mentioned earlier \citep{Church&Church2012, Church_etal2012}.

GX~349+2 is a Sco-like source located approximately 9.2 kpc away \citep{Grimm2002}, with a strong and variable Fe K$\alpha$ line \citep{Cackett2008, Cackett2009, Cackett2010, Cackett2012}. It has been shown to trace out its entire HID in a single day, showing a strong NB and FB, but no HB \citep{diSalvo2001, Iaria_etal2004, Cackett2009}. In this paper we present a spectral analysis of an 80 ks \textit{NuSTAR} observation of GX~349+2, after separating the observation into 5 spectral states: NB, NB/FB vertex (VX), FB1, FB2, and FB3, focusing our analysis on the reflection spectrum of the source in these states.

\begin{figure}[b!]
\includegraphics[trim=1.5cm 1.5cm 2cm 3.0cm, clip=true, width=9cm]{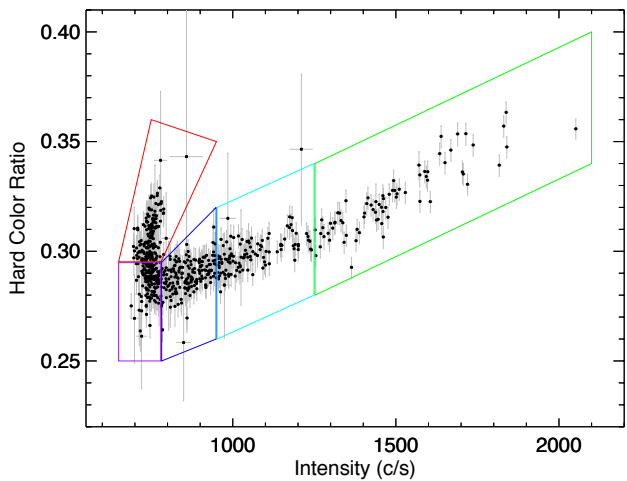}
\caption{Hardness-Intensity Diagram for GX~349+2. The hard color ratio is defined as the 10 -- 16 keV intensity divided by the 6.4 -- 10 keV intensity. Each point represents a 128 s bin. The different spectral shapes are shown, with the normal branch in red (top left), the vertex in purple (bottom left), flaring branch 1 in blue (center-left), flaring branch 2 in cyan (center-right), and flaring branch 3 in green (right). GX~349+2 was not seen in the horizontal branch in our observations, and here the lack of a horizontal branch is quite clear.}
\label{HIDiagram}
\end{figure}

\section{Observation and Data Reduction}\label{sec:data}

The NASA X-ray satellite \textit{NuSTAR} carried out a $\sim$ 80 ks observation of GX~349+2, beginning on June 6, 2016 (ObsID~30201026002). The observation spanned about 150 ks because of Earth occultations, and resulted in a 40 ks exposure time after dead-time corrections. Data was then extracted using the \textsc{nuproducts} tool in 128 second bins, with a 2 arcminute radius to produce light curves for the 3 -- 50 keV range, as well as hard and soft energy light-curves (10 -- 16 and 6.4 -- 10 keV, respectively). The 3 -- 50 keV lightcurve is given in Figure~\ref{LC}, and shows that GX~349+2 was flaring dramatically during the earlier part of the observation, and is generally variable throughout. Since GX~349+2 is a persistently bright source, the background was ignored for the initial purpose of creating the desired lightcurves. \textit{NuSTAR} has two independent CdZnTe crystal detector modules, known as Focal Plane Modules A and B (FPMA and FPMB), and so these separate light curves were added together for each energy band. We then produced an HID of the data, by dividing the hard X-ray light curve by the soft to produce a hard color ratio. In the HID, this hard color ratio is plotted against the broadband (3 -- 50 keV) intensity.

Using the HID, which is shown in Figure~\ref{HIDiagram}, the data was separated into 5 distinct regions, representing 5 different spectral states of GX~349+2. These states are the normal branch (NB), the vertex between the normal and flaring branches (vertex or VX in this paper, often referred to as the soft apex in the literature, as in  \citet{Church&Church2012}), and then the mild, moderate and extreme sections of the flaring branch, or flaring branch 1 (FB1), flaring branch 2 (FB2), and flaring branch 3 (FB3), respectively. The regions were chosen so that each would contain approximately the same number of total counts, however the brightest two regions of the flaring branch (FB2 and FB3) have necessarily fewer counts, as GX~349+2 spends less time in these states. The source covers a very wide range in intensity between these two regions and across the flaring branch in general. Note that a single data point in the flaring branch did not fall closely along the well-defined Z-track, and thus it was not counted in any of these regions. Using dmgti from CIAO Version 4.1, good time interval (or GTI) files were created for each state separately. Data for each spectral state was then extracted again using the \textsc{nuproducts} tool, with a source region of 2 arcminutes centered on the source itself, and a background region of the same size chosen to be well away from the source. The resulting spectra were then used for analysis.

\begin{figure}[t!]
\includegraphics[trim=4.4cm 3.2cm 13.5cm 3.0cm, clip=true, width=9cm]{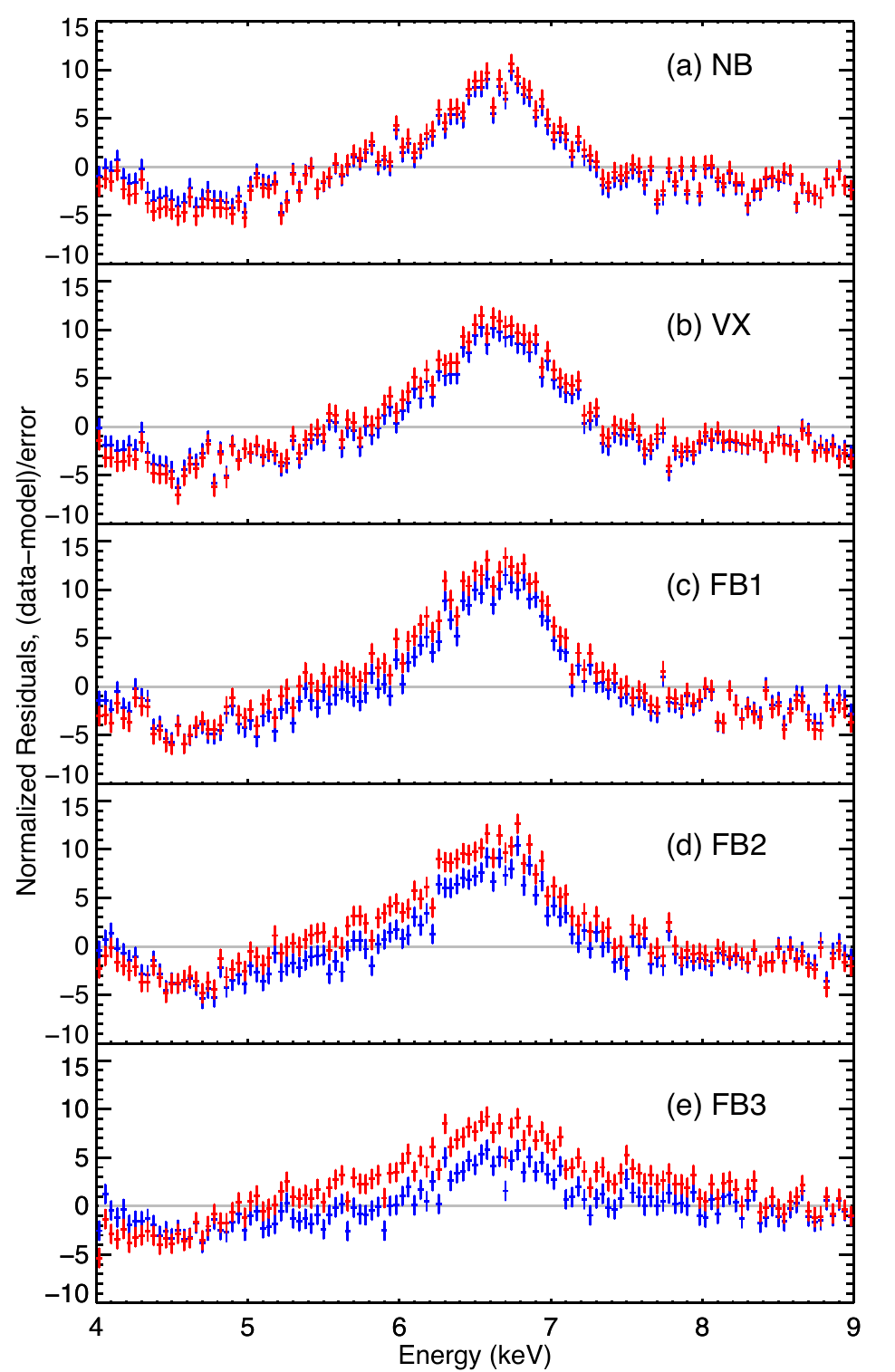}
\caption{Broadened Fe K$\alpha$ line in each spectral state. Each panel represents a different state, with the NB, VX, FB1, FB2, and FB3 shown in panels a, b, c, d, and e, respectively. The y-axis shows the residuals for each spectrum against its best fit continuum model, divided by the error. The best fit continuum model in each state was defined by fitting the entire spectrum. The blue spectrum shows the ratio against the \textsc{diskbb}$+$\textsc{bbodyrad} model, while the red spectrum shows the ratio against the Comptonization continuum model, \textsc{nthcomp}$+$\textsc{bbodyrad}. There are only very small differences in the shape of the line caused by using different continuum models. There is little change in the shape of the line between states, except in the case of FB3, where the line is noticeably shorter and broader.}
\label{FeLine2}
\end{figure}

\section{Analysis and Results}\label{sec:results}

Spectral fitting was carried out using XSPEC version 12.9.1 \citep{Xspec} for each of the five averaged spectra representing the different spectral states. In the NB and VX, the background dominates above $\sim$ 30 keV, while this is only the case in FB1 above 35 keV, and in FB2 and FB3 above 40 keV. Therefore, the spectra were fitted over the energy ranges 3 -- 30 keV, 3 -- 35 keV, and 3 -- 40 keV (NB and VX, FB1, and FB2 and FB3, respectively). Interstellar absorption was taken into account for all spectra using the model \textsc{tbabs} \citep{tbabs}, with the neutral Hydrogen column density fixed to $\textit{N}_{\rm H} = 0.5 \times 10^{22}$ cm$^{-2}$, a value taken from \citet{Kalberla2005}. After fitting the spectra in each region, we tested whether fixing this value had an effect on any of our models, and in each case it either did not change the fit or tended to zero, and we therefore only present results using the fixed value throughout the rest of this paper. Errors are given at the $90\%$ confidence level unless otherwise noted. For each state, the spectra for FPMA and FPMB were originally modeled separately, but with their parameters set to vary only by a constant, so that any notable differences due to an issue with either detector would be easily identifiable. This was done by including the model \textsc{constant} and fixing it equal to 1.0 for FPMA, while allowing the factor to vary for FPMB. After initially fitting the FPMA and FPMB spectra separately for each state, and with the \textsc{constant} factor never varying from unity by more than 0.3 percent, we combined the FPMA and FPMB spectra using \textsc{addascaspec} and will present the results of fitting the combined spectra only.

\subsection{Continuum Modeling}

\begin{deluxetable*}{lllllll}[h!]
\tablewidth{0pt}
\tablecolumns{7}
\tablecaption{Continuum Model Parameters with \textsc{diskbb}}
\tablehead{Component & Parameter & Normal Br. & Vertex & Flaring Br. 1 & Flaring Br. 2 & Flaring Br. 3}
\startdata
\textsc{tbabs} & $\textit{N}_{\rm H}$ ($10^{22}$ $cm^{-2}$)$^{\ast}$ & 0.50 & 0.50 & 0.50 & 0.50 & 0.50
\\
\textsc{bbodyrad} & $kT$ & $2.76 \pm0.02$ & $2.96 \pm0.03$ & $3.07 \pm0.03$ & $3.37 \pm0.05$ & $4.38^{+0.21}_{-0.18}$
\\
& norm & $5.23^{+0.26}_{-0.25}$ & $2.60 \pm0.17$ & $1.97^{+0.14}_{-0.13}$ & $0.99^{+0.12}_{-0.11}$ & $0.11^{+0.04}_{-0.03}$
\\
\textsc{diskbb} & $T_{in}$ & $1.91 \pm0.01$ & $2.02 \pm0.01$ & $2.14 \pm0.01$ & $2.34 \pm0.01$ & $2.69 \pm0.01$
\\
& norm & $56.2^{+1.0}_{-0.9}$ & $48.1 \pm0.7$ & $44.4 \pm0.5$ & $37.7 \pm0.4$ & $30.6 \pm0.2$
\\
\hline
$\chi^2$ (dof) & & 2865.25 (609) & 3615.28 (594) & 4802.64 (622) & 4035.10 (604) & 3614.48 (599)
\\
Reduced $\chi^2$ & & 4.705 & 6.086 & 7.721 & 6.681 & 6.034
\enddata
\tablecomments{Results using a blackbody and thermal disk model for the full spectra. The neutral hydrogen column density is taken from \citet{Kalberla2005}. The temperatures for \textsc{bbodyrad} and \textsc{diskbb} are in keV, and all errors are given at 90 percent confidence intervals. $^{\ast}$~denotes fixed parameters}
\end{deluxetable*}

\begin{deluxetable*}{lllllll}[h!]
\tablewidth{0pt}
\tablecolumns{7}
\tablecaption{Continuum Model Parameters with \textsc{nthcomp}}
\tablehead{Component & Parameter & Normal Br. & Vertex & Flaring Br. 1 & Flaring Br. 2 & Flaring Br. 3}
\startdata
\textsc{tbabs} & $\textit{N}_{\rm H}$ ($10^{22}$ $cm^{-2}$)$^{\ast}$ & 0.50 & 0.50 & 0.50 & 0.50 & 0.50
\\
\textsc{bbodyrad} & $kT$ & $1.35 \pm0.01$ & $1.39 \pm0.01$ & $1.40 \pm0.01$ & $1.46 \pm0.01$ & $1.58 \pm0.01$
\\
& norm & $126.3^{+4.0}_{-3.8}$ & $140.3^{+2.7}_{-3.3}$ & $163.3^{+3.4}_{-3.3}$ & $173.6^{+3.8}_{-2.0}$ & $182.4^{+4.5}_{-2.2}$
\\
\textsc{nthcomp} & Gamma & $1.91 \pm0.01$ & $1.95 \pm0.01$ & $1.88 \pm0.01$ & $1.81 \pm0.01$ & $1.70 \pm0.02$
\\
& $kT_{e}$ & $2.90 \pm0.02$ & $2.97 \pm0.03$ & $2.88 \pm0.02$ & $2.88^{+0.03}_{-0.02}$ & $2.92 \pm0.03$
\\
& $kT_{bb}$ & $0.020^{+0.300}_{-0.020}$ & $0.007^{+0.304}_{-0.007}$ & $0.007^{+0.348}_{-0.007}$ & $0.007^{+0.338}_{-0.007}$ & $0.077^{+0.335}_{-0.077}$
\\
& inp type$^{\ast}$ & $1.0$ & $1.0$ & $1.0$ & $1.0$ & $1.0$
\\
& Redshift$^{\ast}$ & $0.0$ & $0.0$ & $0.0$ & $0.0$ & $0.0$
\\
& norm & $3.10 \pm0.07$ & $3.06 \pm0.07$ & $2.86^{+0.07}_{-0.06}$ & $2.81 \pm0.08$ & $2.90 \pm0.09$
\\
\hline
$\chi^2$ (dof) & & 2368.16 (607) & 2813.25 (592) & 3415.13 (620) & 2422.82 (602) & 1358.26 (597)
\\
Reduced $\chi^2$ & & 3.901 & 4.752 & 5.508 & 4.025 & 2.275
\enddata
\tablecomments{Results from fitting the full spectra using a Comptonization model. The neutral hydrogen column density is taken from \citet{Kalberla2005}. The temperatures for \textsc{bbodyrad} and \textsc{nthcomp} are in keV, while Gamma is the power law photon index. All errors are given as 90 percent confidence intervals. $^{\ast}$~denotes fixed parameters}
\end{deluxetable*}

\begin{deluxetable*}{lllllll}
\tablewidth{0pt}
\tablecolumns{7}
\tablecaption{Diskline Model Parameters}
\tablehead{
Component & Parameter & Normal Br. & Vertex & Flaring Br. 1 & Flaring Br. 2 & Flaring Br. 3}
\startdata
\textsc{tbabs} & $\textit{N}_{\rm H}$ ($10^{22}$ $cm^{-2}$)$^{\ast}$ & 0.50 & 0.50 & 0.50 & 0.50 & 0.50
\\
\textsc{bbodyrad} & $kT$ & $2.55 \pm0.02$ & $2.62 \pm0.02$ & $2.67 \pm0.02$ & $2.80 \pm0.04$ & $3.22^{+0.12}_{-0.09}$
\\
& norm & $10.21^{+0.46}_{-0.45}$ & $7.20^{+0.44}_{-0.41}$ & $6.84^{+0.47}_{-0.46}$ & $5.42^{+0.65}_{-0.62}$ & $1.80^{+0.61}_{-0.51}$
\\
\textsc{diskbb} & $T_{in}$ & $1.71 \pm0.02$ & $1.80^{+0.01}_{-0.02}$ & $1.91 \pm0.02$ & $2.12 \pm0.02$ & $2.57 \pm0.03$
\\
& norm & $82.3^{+2.5}_{-2.3}$ & $69.5^{+2.0}_{-1.8}$ & $62.5 \pm1.6$ & $50.2^{+1.6}_{-1.5}$ & $34.2^{+1.1}_{-1.0}$
\\
\textsc{diskline} & Line E & $6.48 \pm0.03$ & $6.50 \pm0.02$ & $6.47 \pm0.02$ & $6.44 \pm0.03$ & $6.40^{+0.04}_{}$
\\
& $\beta$ & $2.11 \pm0.05$ & $2.07 \pm0.04$ & $2.11 \pm0.04$ & $2.18 \pm0.04$ & $2.45^{+0.03}_{-0.04}$
\\
& $\textit{R}_{in}$ & $6.00^{+0.24}_{}$ & $6.00^{+0.11}_{}$ & $6.00^{+0.18}_{}$ & $6.00^{+0.07}_{}$ & $6.00^{+0.04}_{}$
\\
& ${\textit{R}_{out}}^{\ast}$ & 1000 & 1000 & 1000 & 1000 & 1000
\\
& Incl. & $90.0^{}_{-6.8}$ & $90.0^{}_{-4.8}$ & $90.0^{}_{-6.3}$ & $90.0^{}_{-4.0}$ & $90.0^{}_{-3.1}$
\\
& norm$\times10^{-2}$ & $1.77 \pm0.12$ & $2.07 \pm0.11$ & $2.43 \pm0.11$ & $3.14 \pm0.15$ & $4.93 \pm0.17$
\\
\hline
$\chi^2$ (dof) & & $893.77 (604)$ & $759.28 (589)$ & $859.98 (617) $ & $843.29 (599)$ & $914.63 (594)$
\\
Reduced $\chi^2$ & & 1.480 & 1.289 & 1.394 & 1.408 & 1.540
\enddata
\tablecomments{A \textsc{diskline} component was added to the continuum model, and a new best fit was found including the new parameters. The neutral hydrogen column density is taken from \citet{Kalberla2005}. The energies $kT$ from \textsc{bbodyrad} and $T_{in}$ from \textsc{diskbb} are given in keV, as is the line energy for \textsc{diskline}. $\beta$ gives the powerlaw dependence of disk emissivity on radius, so emissivity goes as $R^{-\beta}$. Inner and outer disk radii are given in units of gravitational radii, $GM/c^2$. When errors are not given for a variable, the parameter is found at the hard upper or lower limit of the model, as was the case for each of the states at $R_{in} = 6.00$ and Incl. $= 90.0$. The line energy in \textsc{diskline} was limited to the range 6.40 -- 6.97~keV. Errors are given for 90 percent confidence intervals. $^{\ast}$~denotes fixed parameters}
\end{deluxetable*}

Initial modeling was done to fit the broad continuum, and a variety of two component models were attempted. First, we used the model \textsc{bbodyrad} in XSPEC to represent the neutron star or boundary region, along with a multicolor disk component, \textsc{diskbb}. These models provided a poor fit, but gave a good base with which to begin more detailed modeling, and our results can be seen in Table~1. The residuals were dominated by a broad Fe K$\alpha$ emission line. A host of other models including the blackbody but also a Comptonization component were also tried in the hopes of finding a better fit to the continuum. A power law spectrum with a high energy cutoff, \textsc{cutoffpl} in XSPEC, has widely been used to model Comptonization in NS LMXB spectra, but this also gave a poor fit. More physical Comptonization models, including \textsc{comptt} \citep{compTT} and \textsc{nthcomp} \citep{nthComp1, nthComp2}, were used as well with a blackbody, and these provided much better statistics. For the \textsc{nthcomp} model, the input parameter `inp~type' was set equal to 1.0, which represents seed photons from a disk that are then up scattered by the higher temperature Comptonizing corona, and the redshift was set equal to zero. The best-fit parameters using \textsc{bbodyrad}$+$\textsc{nthcomp} are given in Table~2. For any of these models, the \textsc{cutoffpl} or Comptonization spectrum closely resembled the hot blackbody spectrum from our earlier model that had included \textsc{diskbb}, and we find noticeably lower temperatures for the blackbody as a result. The asymmetric broadened iron line continued to stand out in the residuals for each of the continuum models we used. The normalized residuals are plotted in Figure~\ref{FeLine2}, which shows the broad Fe line in all 5 states for two of our continuum models, one using \textsc{diskbb} (Table~1), the other utilizing \textsc{nthcomp} (Table~2). It can be seen that the shape of the line is not markedly changed by the continuum model used, and this was true for all of the models we attempted. One noticeable pattern is that, for each choice of continuum model, the thermal components \textsc{bbodyrad} and \textsc{diskbb} showed a general trend of increasing temperature from the NB through the flaring branch.

As an initial attempt to model the broad Fe emission, we added the \textsc{diskline} model component in XSPEC to the \textsc{bbodyrad+diskbb} model. In this case, we modeled the complete spectrum. These results are given in Table~3. Though spectral fits were poor for most of these states, it marks a significant improvement on the full continuum models from before --- the chi-squared improves by greater than 1900 in all cases for a change of 5 degrees of freedom. This model also provides a better fit than any of the two component continuum models in each of the 5 spectral states. However, the best-fit \textsc{diskline} model parameters are clearly an unreliable description of the physical properties of the NS and the accretion disk. In particular, the inclination found a best fit at 90 degrees in every state, most likely as an attempt to match the significant reflection continuum in addition to the emission line. The inner disk radius was pinned at the lower limit of the model, at 6.0 $R_g$, for each of the 5 states as well. A more realistic model is clearly desired, not simply in order to improve the spectral fits, but also to take into account reflection off of the accretion disk, which produces not just the Fe line but also other reflection features, such as a Compton backscattering hump at higher energies. As an additional test, we also attempted to fit the Fe emission with the inclusion of several narrow Gaussian lines representing Fe~XXV and Fe~XXVI, but had poor results modeling the broad excess around 6 -- 7~keV. Though some previous studies on Z sources have not required a reflection component \citep{Church_etal2012}, the detection of a broad Fe line in our observation, as well as our poor spectral fits, suggests that disk reflection is indeed seen here. This is consistent with previous work on GX~349+2 that has required a reflection component \citep{Cackett2009, Iaria2009, Cackett2010, Cackett2012}.

\subsection{Reflection Modeling}

Our first look at modeling the broad Fe line with a reflection spectrum used a model where the incident X-rays producing reflection are emitted by a blackbody spectrum. In this case, it is the NS or boundary layer that provides this flux. For this model we used the \textsc{bbodyrad+diskbb} continuum. For the reflection component we utilized a modified version of \textsc{reflionx} by \citet{Ross&Fabian2005} that models reflected emission off an accretion disk from a blackbody rather than a power law \footnote{http://www-xray.ast.cam.ac.uk/$\sim$mlparker}. We will call this model \textsc{reflionx$_{BB}$}. Its parameters include the ionization parameter $\xi$, the Fe abundance measured relative to the solar abundance, the temperature of the seed photons producing the reflection, the redshift $z$, and a normalization. We fixed the Fe abundance equal to 1, since from initial modeling it was poorly constrained. The blackbody temperature was tied between the reflection and continuum components, since the boundary region very near the NS is widely considered to be the source of the hard X-rays that produce reflection \citep[e.g.][]{Cackett2010, DAi2010}. Finally, the redshift was set and fixed to $z=0$, as should be the case for any Galactic source.

To take into account the relativistic effects in the strong gravity environment near the neutron star, we used \textsc{relconv} \citep{Dauser2010} convolved with the reflected emission. Although \textsc{relconv} is able to model a broken power law for emissivity throughout the disk at different radii, we fixed the emissivity index to have one value for simplicity, to the commonly used value 3.0 \citep[see][]{Wilkins&Fabian2012}. Other parameters include the NS or black hole spin, the disk inclination, the inner and outer disk radii, and a parameter for limb darkening or brightening. 

\begin{figure*} [t!]
\centering
\includegraphics[trim=0.25cm 2.65cm 2cm 1.5cm, clip, width=9cm]{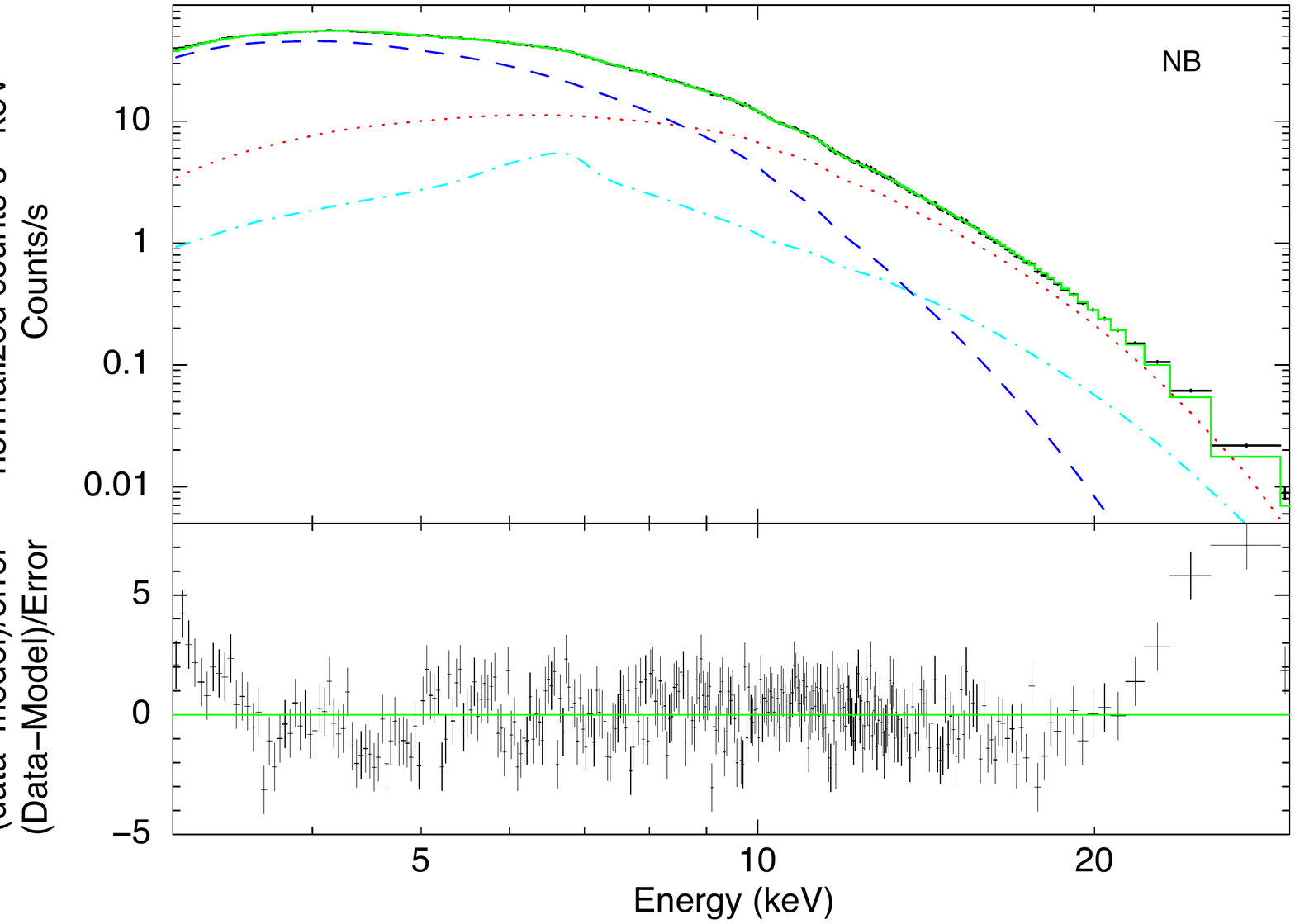}\includegraphics[trim=1.5cm 2.65cm 2cm 1.5cm, clip=true, width=8.5cm]{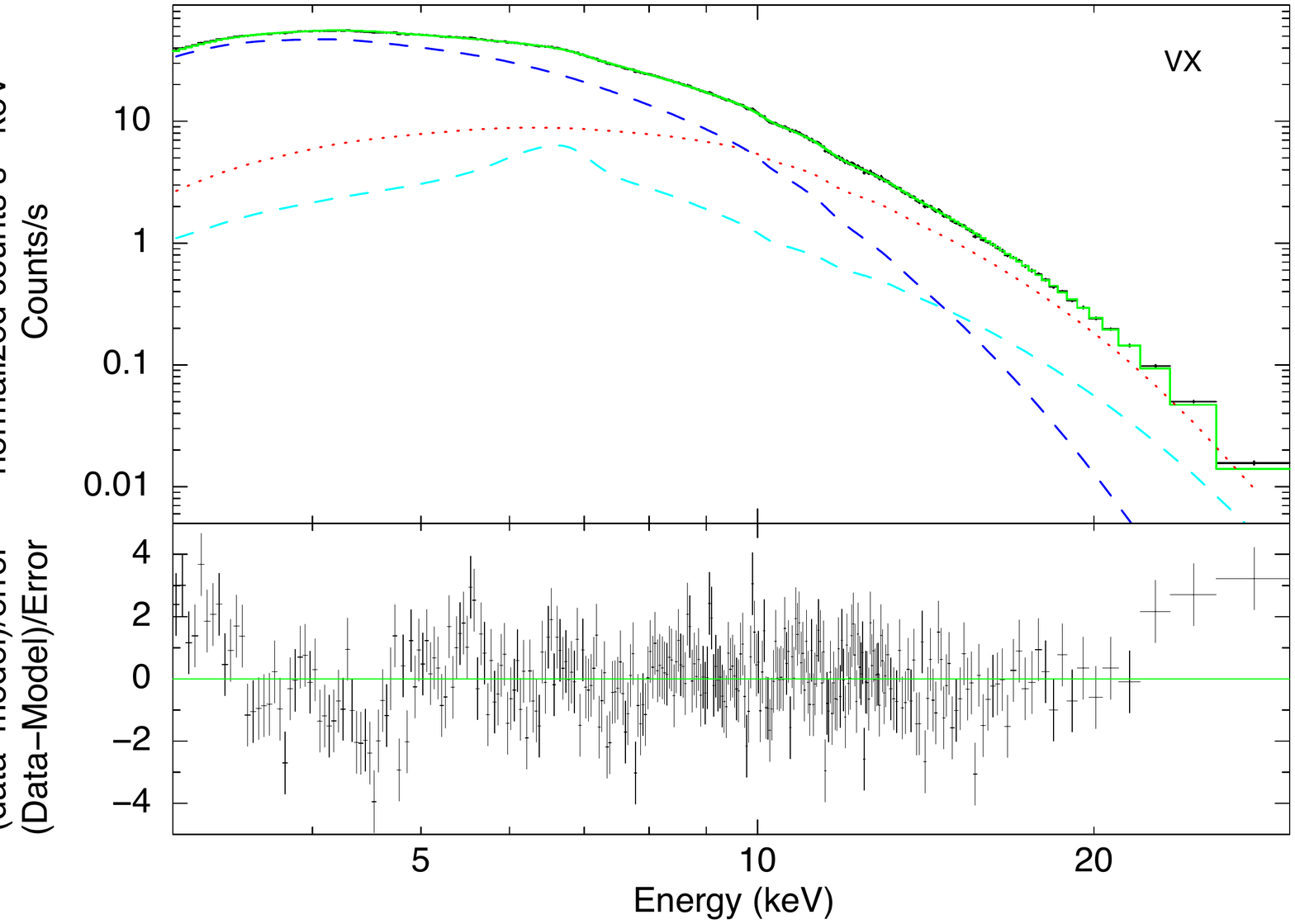}
\includegraphics[trim=0.25cm 1cm 2cm 2.2cm, clip=true, width=9cm]{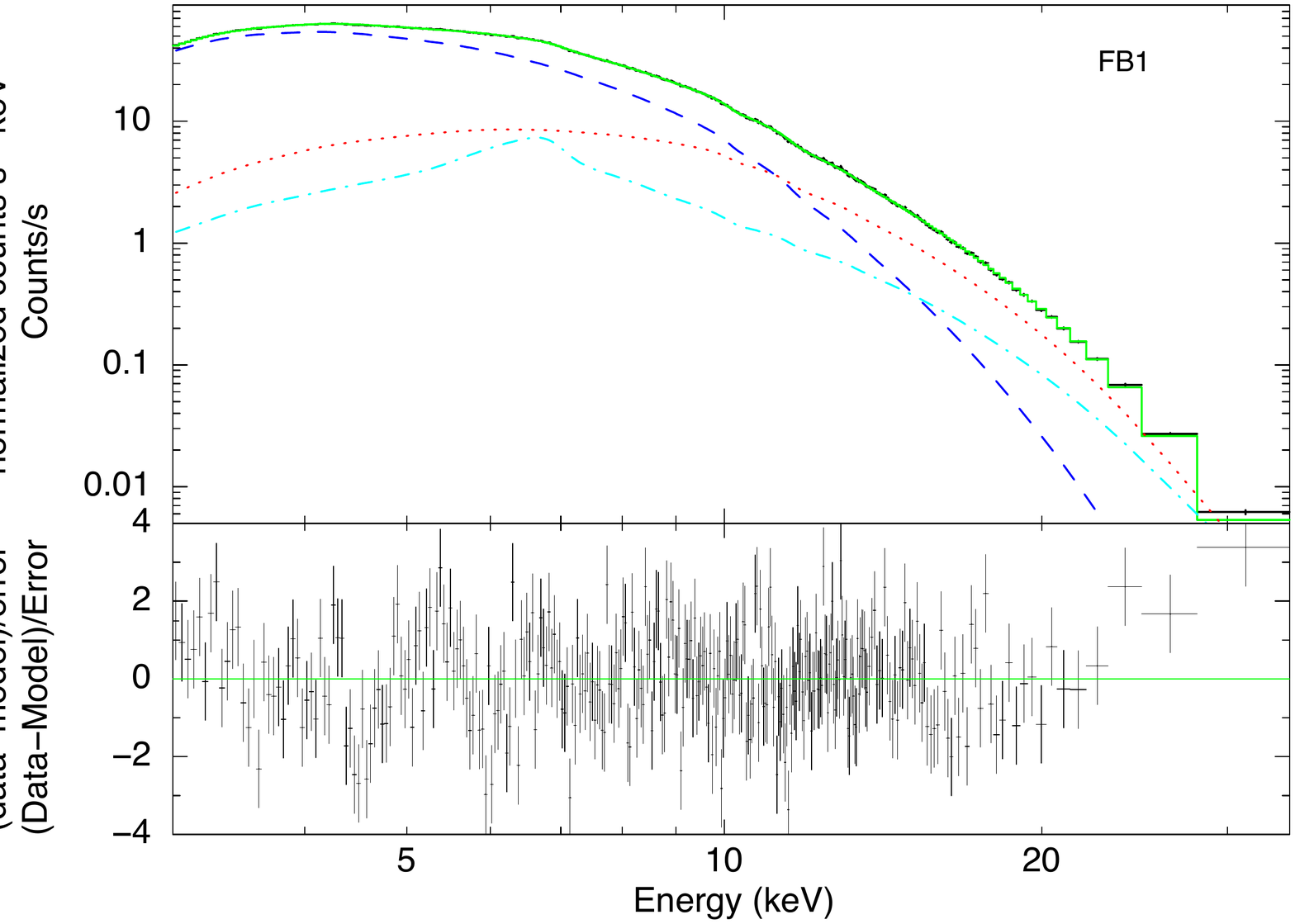}\includegraphics[trim=1.5cm 1cm 2cm 2.2cm, clip=true, width=8.5cm]{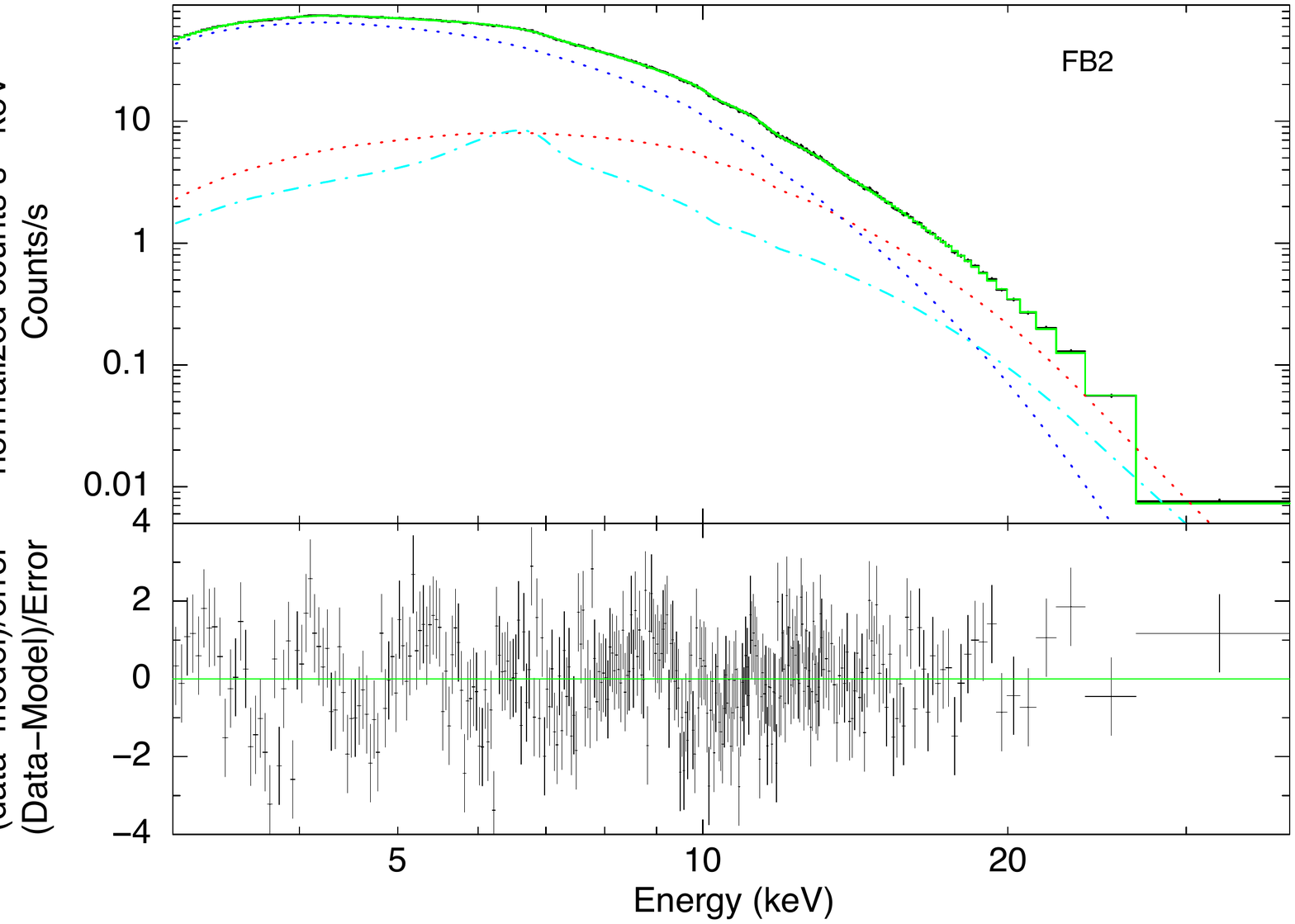}
\caption{Spectra for the NB through FB2, plotted with additive model components. The combined FPMA and FPMB data, added using \textsc{addascaspec}, is shown in black. The best-fit model is the solid green line, whereas the additive model components are shown as follows: \textsc{bbodyrad} is the red dotted line, \textsc{diskbb} is the blue dashed line, and \textsc{reflionx$_{BB}$} is the cyan dot-dashed line. The lower panel of each plot shows the difference between the data and the model, normalized by the square root of the number of counts in each energy bin.}
\label{spectra}
\end{figure*}

\begin{figure}
\centering
\includegraphics[trim=0.25cm 1cm 2cm 1.5cm, clip=true, width=9cm]{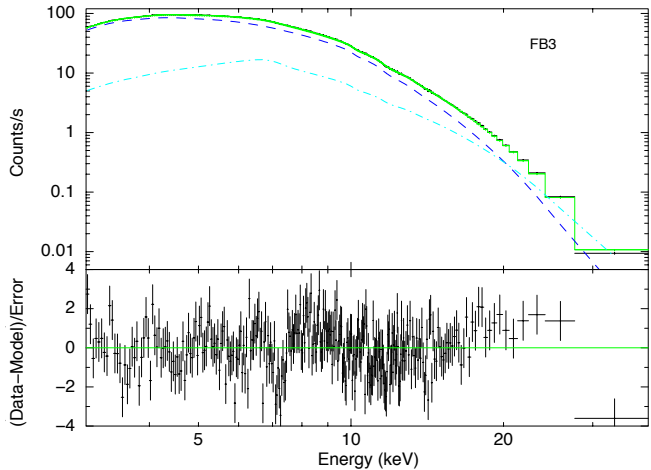}
\caption{Flaring branch 3 spectrum with additive model components. As with the other plotted spectra, the added FPMA and FPMB data is given in black. The solid green line represents the best-fit model, while the dashed blue line is the \textsc{diskbb} component, and the dot-dashed cyan line shows the \textsc{reflionx$_{BB}$} model. The normalized difference between the data and the model is shown in the bottom panel. This best fit model is unrealistic due to the almost zero blackbody normalization, and as can be seen in the figure, the dotted red blackbody component seen in the other models does not appear here.}
\label{FB3}
\end{figure}

For the NS spin, although the spin of GX~349+2 is unknown, we assume that it is similar to other known NS LMXBs with a weakly magnetized neutron star and spin parameter $a \lesssim 0.3$. We therefore set the \textsc{relconv} spin parameter value $a = 0$ \citep{Galloway_etal2008, Miller&Miller2015}. Furthermore, the general relativistic effects of the slow NS spin should have little effect on the inner disk radius. For instance, $a = 0.3$ has $R_{ISCO} = 5.0$ $R_g$ compared to $R_{ISCO} = 6.0$ $R_g$ for $a = 0$. The outer disk radius was fixed to 1000 $R_g$, while the inner disk radius and inclination were allowed to vary. Finally, the limb darkening parameter was set to 0 to treat the source as isotropic. The results from relativistic reflection modeling are given in Table~4, and in Figures~\ref{spectra} and \ref{FB3} we show the modeled spectra for each of the states along with normalized residuals, given as the difference between the data and the model divided by the error.

Initial fitting of the VX and FB2 spectra gave some surprising results, namely that in both states the best-fit inclination dropped to below 10 degrees. In the VX, the best-fit ionization parameter jumped to over 600, while it was below 200 for both the neighboring states: the NB and FB1. The normalization of the \textsc{reflionx$_{BB}$} model dropped in both states as well, when compared to the fits of the other regions. After some investigation, it became clear that the ionization parameter and normalization were degenerate, especially in the case of the vertex spectra, and the high ionization and low normalization for \textsc{reflionx$_{BB}$} were causing an unrealistic fit for the inclination. In response, we limited the range of the inclination to within the $90\%$ confidence range of the weighted average from the NB and FB1, which was $24.42^{+0.91}_{-1.21}$ degrees. With this limitation, we found more reasonable \textsc{reflionx$_{BB}$} parameters with only a small change in the chi-squared, and these are the values that we report in Table~4. We also found an unrealistic inclination for FB3, which in the best-fit model had an inclination of 89.0 degrees, and so we limited the FB3 inclination in the same way. This produced a new best fit where the \textsc{bbodyrad} normalization dropped to near zero, as can be seen in Table~4, which will be discussed in \S 3.3.

As would be expected for a Sco-like Z source \citep{Church_etal2012}, the blackbody temperature we measure is very high for a NS LMXB, from $2.45^{+0.02}_{-0.01}$ keV in the NB up to $2.91^{+0.04}_{-0.18}$ keV in the extreme flaring branch, FB3. The accretion disk temperature also increases from the NB through the vertex and along the flaring branch, from $1.65^{+0.02}_{-0.01}$ keV to $2.52^{+0.02}_{-0.15}$ keV. The blackbody and disk normalizations, on the other hand, both show a decreasing trend from the NB to FB2, while in FB3 the blackbody normalization drops to almost nothing, with a $90\%$ upper limit of 3.69 and a best-fit value of 0.00002. This near-zero value is not expected in the scenario of unstable nuclear burning, though we do not interpret it as evidence against such burning. The disk normalization also drops in FB3, but following the pattern set from the NB to FB2 and without approaching zero.

\subsection{Investigating Flaring Branch 3}

\begin{figure}[h!]
\centering
\includegraphics[trim=3.2cm 2.8cm 11cm 5cm, clip=true, width=9cm]{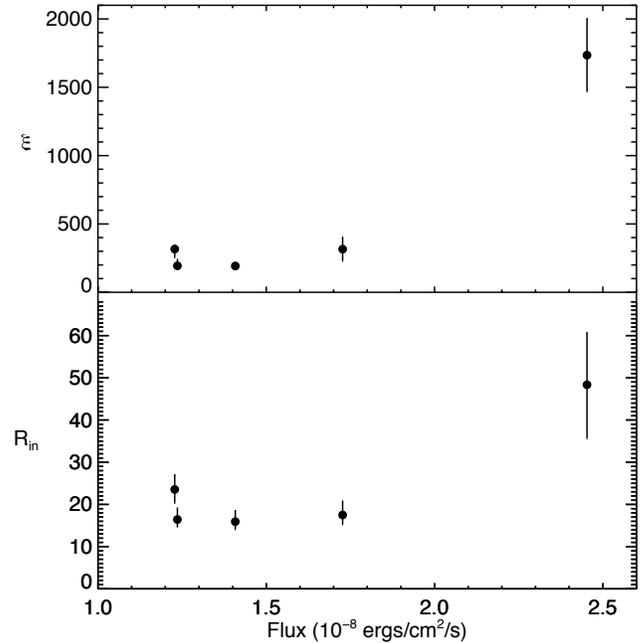}
\caption{Reflection results and flux. The states from lower to higher flux are ordered VX, then NB, and then FB1 through FB3. $\xi$, the ionization parameter for the disk, and the inner disk radius $R_{in}$ are weakly correlated and poorly constrained during flaring. While it may appear as though the inner disk radius increases during extreme flaring, the measurements for FB3 should not be considered reliable. Errors shown are $90\%$ confidence values.}
\label{radius}
\end{figure}

In the reflection spectrum, some notable differences occurred between FB3 and the other spectral states. Figure~\ref{radius} shows the ionization parameter and inner radius of the disk as a function of the source's flux. From the NB through the moderate flaring branch FB2, the inner disk radius is roughly consistent with remaining unchanged, varying from 15.9 $R_g$ in FB1 to 23.5 $R_g$ for the VX. In fact, the inner disk radius measurement is very similar for the NB, FB1, and FB2. In FB3, however, the inner disk radius is dramatically higher and poorly constrained, at $48.3^{+12.4}_{-12.6} R_g$. This measurement would support a change in the geometry of the accretion disk during flaring, if it were to be taken at face value. However, this is most likely an unrealistic measure of the accretion disk, which will be discussed later. Another dramatic change in the reflection spectrum was the ionization parameter $\xi$, which is nearly constant around $\sim$~200 or 300, except when it rises to $1735^{+270}_{-264}$ in FB3 (see Fig.~\ref{radius}). The inclination for FB3, on the other hand, agrees well with the values found for the NB and in FB1, and these are consistent with previous measurements of the inclination of GX~349+2, which range from 18 to 43 degrees \citep{Iaria2009, Cackett2010}.

The extreme flaring state FB3 proved difficult to model with any physically acceptable scenario. A near-zero normalization for the blackbody component in FB3 would produce a physical inconsistency, since it is the blackbody that provides the source of hard X-rays required for reflection to occur. While a non-zero normalization was found in one fit, it was coincident with an impossible inclination value, and with the inclination limited to a realistic value the blackbody normalization dropped drastically. A broad Fe K$\alpha$ line is undoubtedly seen, as has already been discussed and shown in Figure~\ref{FeLine2}, which is a signature of reflection off of the accretion disk. This provided further motivation, along with our initial continuum modeling, to study GX~349+2 with a different reflection model, which is discussed in \S 3.4.

The issues with modeling the reflection spectrum self-consistently in FB3 cast doubt on how reliable the inner disk radius measurement is in this state. One possible complexity in modeling this high flux state is the degeneracy between the ionization parameter and the inner disk radius. As the ionization parameter increases, the effects of Compton scattering on line broadening become more important. Thus, a higher ionization parameter is compensated by a larger inner disk radius. This seems to be driving the fit to a high ionization parameter. In conclusion, we do not have a satisfactory (self-consistent) reflection fit for FB3 with a trustworthy inner disk radius measurement.

To study FB3 in more detail, and considering GX~349+2 has many distinct flares throughout the first half of our observation, we extracted spectra for each of these individual flares to model independently. We chose five individual flares, counting only those that rose above 1250 counts/s and would thus be considered part of FB3. Fitting these flares individually, we found that in four out of the five flares, the \textsc{bbodyrad} normalization did not drop to zero. Furthermore, the other continuum parameters varied drastically in between flares --- for instance, the blackbody temperature ranged from 2.34 to 2.93 keV, while the \textsc{diskbb} temperature varied from 1.49 to 2.51 keV. This is a broader range of spectral parameters than can be seen across any of the five distinct spectral states. The differences between the individual flare spectra can be seen visually in Figure~\ref{flares} as well. These differences are likely to explain at least some of the difficulty in modeling FB3 with a physically consistent reflection model.

\begin{deluxetable*}{lllllll}
\tablewidth{0pt}
\tablecolumns{7}
\tablecaption{Reflection Model Parameters -- \textsc{reflionx$_{BB}$}}
\tablehead{
Component & Parameter & Normal Br. & Vertex & Flaring Br. 1 & Flaring Br. 2 & Flaring Br. 3}
\startdata
\textsc{tbabs} & $\textit{N}_{\rm H}$ ($10^{22}$ $cm^{-2}$)$^{\ast}$ & 0.50 & 0.50 & 0.50 & 0.50 & 0.50
\\
\textsc{bbodyrad} & $kT$ & $2.45^{+0.02}_{-0.01}$ & $2.49^{+0.01}_{-0.02}$ & $2.50^{+0.02}_{-0.01}$ & $2.63 \pm0.03$ & $2.91^{+0.04}_{-0.18}$
\\
& norm & $11.08^{+0.34}_{-0.52}$ & $8.36^{+0.43}_{-0.35}$ & $8.02^{+0.44}_{-0.58}$ & $6.53^{+1.00}_{-1.03}$ & $0.00002^{+3.69}_{}$
\\
\textsc{diskbb} & $T_{in}$ & $1.65^{+0.02}_{-0.01}$ & $1.73 \pm0.01$ & $1.85 \pm0.02$ & $2.06 \pm0.03$ & $2.52^{+0.02}_{-0.15}$
\\
& norm & $91.0^{+2.4}_{-3.4}$ & $78.6 \pm1.8$ & $68.6^{+1.7}_{-2.3}$ & $53.7^{+2.7}_{-2.6}$ & $33.7^{+5.5}_{-0.2}$
\\
\textsc{relconv} & Index$^{\ast}$ & 3.00 & 3.00 & 3.00 & 3.00 & 3.00
\\
& $a^{\ast}$ & 0.0 & 0.0 & 0.0 & 0.0 & 0.0
\\
& Incl. & $24.6^{+1.5}_{-1.8}$ & $23.2^{+1.1}_{}$ & $24.3^{+1.1}_{-1.6}$ & $23.2^{+1.3}_{}$ & $25.1$
\\
& $\textit{R}_{in}$ & $16.4^{+2.8}_{-1.8}$ & $23.5^{+3.6}_{-3.2}$& $15.9^{+2.7}_{-1.7}$ & $17.5^{+3.4}_{-2.2}$ & $48.3^{+12.4}_{-12.6}$
\\
& ${\textit{R}_{out}}^{\ast}$ & $1000.0$ & $1000.0$ & 1000.0 & 1000.0 & 1000
\\
& limb$^{\ast}$ & 0.0 & 0.0 & 0.0 & 0.0 & 0.0
\\
\textsc{reflionx}$_{BB}$ & $\xi$ & $194^{+49}_{-15}$ & $316^{+32}_{-60}$ & $193^{+19}_{-20}$ & $316^{+89}_{-84}$ & $1735^{+270}_{-264}$
\\
& Fe Abund.$^{\ast}$ & 1.00 & 1.00 & 1.00 & 1.00 & 1.00
\\
& $kT$ & \multicolumn{5}{c}{Equal to \textsc{bbodyrad} $kT$} 
\\
& z$^{\ast}$ & 0.0 & 0.0 & 0.0 & 0.0 & 0.0
\\
& norm & $3.53^{+0.40}_{-0.80}$ & $2.34^{+0.61}_{-0.17}$ & $4.84^{+0.78}_{-0.53}$ & $3.31^{+1.36}_{-0.54}$ & $2.60^{+0.06}_{-0.12}$
\\
Unabsorbed Flux, 3-30 keV & (10$^{-8}$ ergs/cm$^2$/s) & $1.236 \pm0.001$ & $1.228 \pm0.001$ & $1.408 \pm0.001$ & $1.727 \pm0.001$ & $2.453 \pm0.002$
\\
$L/L_{\rm Edd}$ & & $0.710$ & $0.705$ & $0.808$ & $0.991$ & $1.409$
\\
\hline
$\chi^2$ (dof) & & $849.49 (605)$ & $687.81 (590)$ & $729.23 (618)$ & $707.84 (600)$ & $740.68 (595)$
\\
Reduced $\chi^2$ & & 1.404 & 1.166 & 1.180 & 1.180 & 1.245
\enddata
\tablecomments{The neutral hydrogen column density for \textsc{tbabs} is taken from \citet{Kalberla2005}. Temperatures for \textsc{bbodyrad} and \textsc{diskbb} are given in keV. The Index parameter for \textsc{relconv} is the disk emissivity index, giving the powerlaw dependence of emissivity on radius. $a$ is the dimensionless spin parameter for the NS, and the inner and outer disk radii were converted to gravitational radii, $GM/c^2$, for a non-rotating NS or black hole. The limb parameter is set to 0 to avoid limb brightening or darkening. $\xi$ is the ionization parameter for the \textsc{reflionx$_{BB}$} model. Luminosities were calculated assuming a distance of 9.2 kpc, for spherical accretion around a 1.4 $M_\odot$ NS. Inclination for the VX and FB2 were limited to the range 23.21 -- 25.33 degrees. Errors are given to show 90~percent confidence intervals. $^{\ast}$~denotes fixed parameters}
\end{deluxetable*}

\begin{deluxetable*}{lllllll}
\tablewidth{0pt}
\tablecolumns{7}
\tablecaption{Reflection Model Parameters -- \textsc{relxill}}
\tablehead{
Component & Parameter & Normal Br. & Vertex & Flaring Br. 1 & Flaring Br. 2 & Flaring Br. 3}
\startdata
\textsc{tbabs} & $\textit{N}_{\rm H}$ ($10^{22}$ $cm^{-2}$)$^{\ast}$ & 0.50 & 0.50 & 0.50 & 0.50 & 0.50
\\
\textsc{bbodyrad} & $kT$ & $2.00 \pm0.02$ & $1.78 \pm0.03$ & $1.87 \pm0.01$ & $1.89 \pm0.01$ & $2.03^{+0.01}_{-0.02}$
\\
& norm & $17.0 \pm0.4$ & $28.9 \pm2.1$ & $34.3^{+1.0}_{-0.6}$ & $46.5^{+1.4}_{-1.0}$ & $59.5^{+1.3}_{-1.5}$
\\
\textsc{relxill} & Index$^{\ast}$ & 3.00 & 3.00 & 3.00 & 3.00 & 3.00
\\
& $a^{\ast}$ & 0.0 & 0.0 & 0.0 & 0.0 & 0.0
\\
& Incl. & $35.9^{+0.3}_{-1.0}$ & $36.8 \pm0.6$ & $37.4^{+0.2}_{-0.6}$ & $38.4^{+0.2}_{-0.8}$ & $53.4 \pm4.2$
\\
& $\textit{R}_{in}$ & $6.0^{+0.6}_{}$ & $6.0^{+0.3}_{}$ & $6.0^{+0.3}_{}$ & $6.0^{+0.4}_{}$ & $12.6^{+2.8}_{-3.8}$
\\
& ${\textit{R}_{out}}^{\ast}$ & $1000.0$ & $1000.0$ & 1000.0 & 1000.0 & 1000.0
\\
& z$^{\ast}$ & 0.0 & 0.0 & 0.0 & 0.0 & 0.0
\\
& $\Gamma$ & $1.19 \pm0.01$ & $1.24 \pm0.01$ & $1.17 \pm0.01$ & $1.10 \pm0.02$ & $1.00^{+0.01}_{}$
\\
& log$\xi$ & $3.42 \pm0.04$ & $3.09^{+0.12}_{-0.06}$ & $3.40 \pm0.03$ & $3.36 \pm0.03$ & $3.45^{+0.05}_{-0.06}$
\\
& Fe abund.$^{\ast}$ & 1.0 & 1.0 & 1.0 & 1.0 & 1.0
\\
& Ecut & $5.00^{+0.01}_{}$ & $5.00^{+0.01}_{}$ & $5.00^{+0.01}_{}$ & $5.00^{+0.01}_{}$ & $5.07^{+0.03}_{-0.04}$
\\
& refl. frac. & $0.61^{+0.06}_{-0.05}$ & $0.62^{+0.02}_{-0.03}$ & $0.91^{+0.09}_{-0.07}$ & $0.88^{+0.11}_{-0.08}$ & $1.35^{+0.55}_{-0.22}$
\\
& norm $\times10^{-2}$ & $1.69 \pm0.09$ & $1.91^{+0.06}_{-0.08}$ & $1.47 \pm0.09$ & $1.62 \pm0.13$ & $1.52^{+0.31}_{-0.27}$
\\
\hline
$\chi^2$ (dof) & & $799.53 (604)$ & $741.49 (589)$ & $966.70 (617)$ & $887.94 (599)$ & $964.36 (594)$
\\
Reduced $\chi^2$ & & 1.324 & 1.259 & 1.567 & 1.482 & 1.624
\enddata
\tablecomments{Neutral hydrogen column density is taken from \citet{Kalberla2005}. Temperature for \textsc{bbodyrad} is given in keV. The Index parameter for \textsc{relconv} is the disk emissivity index, giving the powerlaw dependence of emissivity on radius. $a$ is the dimensionless spin parameter for the NS, and the inner and outer disk radii were converted to gravitational radii, $GM/c^2$, for a non-rotating NS or black hole. The limb parameter is set to 0 to avoid limb brightening or darkening. Log$\xi$ relates to the ionization parameter $\xi$ for the \textsc{relxill} model. 90 percent confidence intervals are shown for the errors. $^{\ast}$~denotes fixed parameters}
\end{deluxetable*}

We also dissected FB3 by further separating it into smaller bins of intensity. By dividing the region of FB3 at 1325, 1425, 1500, and 1650 counts/second, we were able to extract 5 spectra consisting of roughly the same number of counts. In this way, we hoped to investigate how the source might change with intensity during intense flaring, and perhaps gain some insight into our difficulty with properly modeling FB3. Fitting these spectra with the \textsc{reflionx$_{BB}$} model, we found that the blackbody component could vary drastically in temperature, from 2.40 to 3.04 keV, and that while higher values caused the blackbody normalization to drop to zero, lower temperatures corresponded to a non-zero normalization. The inner disk radius remained unchanged from the best-fit value for FB3 throughout the 4 spectra of lowest intensity within FB3, while the disk inclination and ionization both varied wildly. Unfortunately, our model revealed no discernible pattern with changes of intensity along FB3, and doing so provided little to no additional insight into the state's difficult spectrum.

\subsection{Alternative Reflection Models}

One of the few other available reflection models that can work for neutron stars as well as black holes assumes incident radiation in the form of a power law with a high energy cutoff. We chose the relatively new model \textsc{relxill} \citep{DauserGarcia_etal2014, GarciaDauser_etal2014}, which provides a more acceptable range of available cutoff energies than most models of this type, to use with a blackbody in addition to the power law continuum. \textsc{relxill} includes relativistic blurring with the same parameters as \textsc{relconv}: the disk emissivity index, NS spin, disk inclination, and the inner and outer disk radii. It also contains an option for redshift, which is again set equal to zero. The other \textsc{relxill} parameters include the power law photon index $\Gamma$ and normalization, the high energy cutoff for the power law emission, the logarithm of the ionization of the disk, log$\xi$, the Fe abundance relative to solar values, and the reflection fraction. This reflection fraction is defined as the ratio of the intrinsic intensity emitted towards the disk compared to the that which escapes to infinity. \textsc{relxill} provides both the incident flux and reflected emission from the disk, and is therefore self-consistent. As with our earlier reflection modeling, we kept the Fe abundance and disk emissivity fixed. The best-fit results for our five different spectra are given in Table~5.

Our best fits with \textsc{relxill} do not fit the spectra as well statistically as our first reflection model using \textsc{reflionx$_{BB}$}, except for the NB, which was the poorest fit for that reflection model and is the best fit using \textsc{relxill}. We once again find a high blackbody temperature across all 5 states, ranging between 1.78 and 2.03 keV. This was not the case during initial continuum fitting with Comptonization models like \textsc{nthcomp}. The blackbody normalization also increases from the NB through the rest of the Z track to FB3, rather than decreasing. The power law index is lower than our initial continuum fitting as well, only dropping to the lower limit of the model at 1.0 in FB3. The model has a hard coded lower limit of 5.0 keV for the high energy cutoff, and in all states excluding FB3 the fit tends toward this hard limit. Such a low high energy cutoff has been seen for NS LMXB sources before, including GX~349+2, and is not too uncommon \citep{Church_etal2014}. The inclination is consistent throughout the first four states, as is the inner disk radius. This value too was found to be at the model's lower limit, which in this case is 1.0 ISCOs, or 6.0 $R_g$ when $a = 0$. This is drastically lower than our previous measurements of the inner disk radius, however it does support the result that the inner disk radius remains unchanged throughout the Z track of GX~349+2. The ionization parameter was found to be well over 1000 in each state as well, which differs from the values found earlier. While the \textsc{relxill} model has no difficulty explaining FB3, it isn't able to provide statistically acceptable fits of all five states and is limited by the ranges of several of the parameters mentioned when fitting the spectra for GX~349+2. As a result, we focus our discussion on the measurements from our earlier reflection modeling using \textsc{reflionx$_{BB}$}.

\section{Discussion}\label{sec:disc}

We have performed spectral analysis of the recent \textit{NuSTAR} observation of the NS LMXB GX~349+2, investigating changes throughout the Z-track. The time-averaged reflection spectrum of GX~349+2 has been studied previously by \citet{Cackett2010}, and attempts to understand the changes in this source's spectrum at different levels of flux have also been made \citep{Iaria_etal2004, Iaria2009}. A general picture of what causes Z sources like GX~349+2 to move about their Z-tracks has been suggested by \citet{Church_etal2012}, after modeling the continuum spectra of GX~349+2, Sco~X-1, and GX~17+2. However, by modeling the broad Fe K line and broadband spectrum with a self-consistent reflection model, a clearer picture might be obtained, adding to our understanding of what causes Z sources to move about on their HIDs.

From initial fitting of the continuum, several patterns became clear. In particular, the temperature of the blackbody component increases steadily from the NB through the end of the flaring branch. This increase in blackbody temperature along the flaring branch has been identified before in Cyg-like Z sources, and suggests unstable nuclear burning on the surface of the NS as the cause of flaring \citep{Church&Church2012}. For Sco-like sources, which flare much more frequently than Cyg-like sources, that increase in blackbody temperature is not seen. Previous measurements of GX~349+2 and Sco~X-1 show that the blackbody component in these sources almost never drops below 2 keV \citep{Church_etal2012}, which we observe in our observation as $kT$ for the \textsc{bbodyrad} component reaches its lowest value at $2.45^{+0.02}_{-0.01}$ keV in the NB, according to our spectral modeling. A high and approximately constant temperature is still very much consistent with unstable nuclear burning, since frequent flaring prevents these sources from ever cooling below 2 keV. Whether or not the interpretation of the blackbody component connects it to a boundary layer or the surface of the neutron star itself, these temperatures are likely to represent unstable nuclear burning during flares in the case of GX~349+2. This increase in blackbody temperature is independent of the models used to fit the continuum emission, however with \textsc{bbodyrad}$+$\textsc{nthcomp} we did not find blackbody temperatures above 2 keV. Other trends in continuum parameters depended on the choice of model used, and so we do not discuss them further.

\begin{figure}[t!]
\centering
\includegraphics[trim=1.0cm 1.1cm 1.5cm 2.5cm, clip=true, width=9cm]{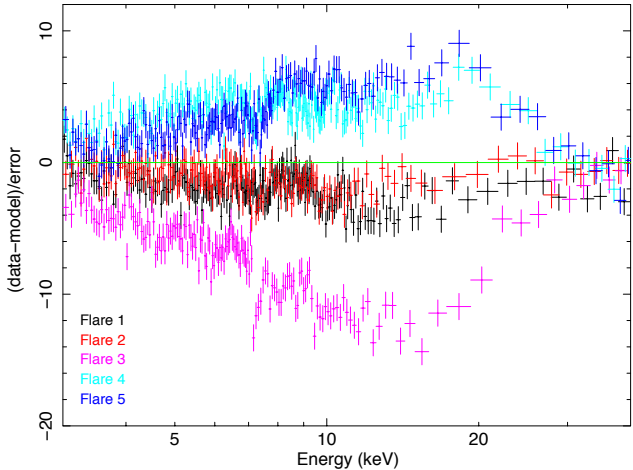}
\caption{Individual flare residuals divided by the errors, as measured against the best-fit FB3 reflection model from Table~4. The spectra for FPMA and FPMB were added together to avoid overcrowding the figure. The spectra of these individual flares are not indistinguishable. Combined, they produce the FB3 spectrum.}
\label{flares}
\end{figure}

From reflection modeling with \textsc{reflionx$_{BB}$}, the NB and FB1 agree strongly on a disk inclination of $\sim$ 25 degrees. Furthermore, the accretion disk appears to be truncated by a boundary layer or perhaps the magnetic field of the neutron star, as the smallest inner disk radius is measured to be $15.9^{+2.7}_{-1.7}$ gravitational radii, corresponding to $33.1^{+5.6}_{-3.5}$ km for a canonical 1.4 $M_\odot$ neutron star. The weighted mean of the NB, VX, FB1 and FB2 inner disk radii gives $R_{in} = 17.5^{+1.5}_{-1.0}$ $R_g$, or 36.4 km. Since this is larger than typical estimates for the radius of such a neutron star, this measurement could be used to set an upper limit on the neutron star radius. Alternately, this inner disk radius measurement can be thought of as evidence of a truncated disk. We can estimate the maximum radius of a boundary layer surrounding the neutron star and truncating the disk, following the work of \citet{Popham&Sunyaev2001}, and in particular by using equation (25) of that work along with an estimated mass accretion rate for GX~349+2. We estimated the bolometric flux of the source to find the mass accretion rate, by calculating the 0.1--200 keV flux of the best-fit model in each state using \textsc{cflux} in XSPEC, and then by calculating the luminosity given a distance of 9.2~kpc and then the mass accretion rate using $L_{acc} = G M \dot{M} / R_{\ast}$ with $M = 1.4$~$M_\odot$ and $R_{\ast} = 10$~km. We find a boundary layer would extend above the surface of the neutron star between 9.1 and 19.4 $R_g$ (19.0 and 40.3 km) from the lowest flux state, the VX, to FB3. If instead the neutron star's magnetic field truncates the disk, we can estimate the strength of such a magnetic field following equation (1) of \citet{Cackett_etal2009}, which is adopted from \citet{Ibragimov&Poutanen2009}. Assuming an accretion efficiency of $\eta = 0.2$ \citep{Sibgatullin2000}, and setting both the angular anisotropy $f_{ang}$ and geometrical conversion factor $k_{A}$ to unity, we can come up with a reasonable estimate for the magnetic field required to truncate the disk. We took the average 0.1--200 keV flux of the first four states, $2.39\times10^{-8}$ ergs/cm$^2$/s, as well as the average inner accretion disk radius for those four states, 17.5 $R_g$. This gives an estimate of the magnetic field strength at the poles of the neutron star; assuming a radius of 10 km for the NS, we find a magnetic field strength of $3.07\times10^9$~G. It should be mentioned that no pulsations have been detected for GX~349+2. Our estimate should only be thought of as an upper limit, in the case that the neutron star's magnetic field truncates the accretion disk.

These results are comparable to a recent analysis of the reflection spectrum of another Z source, Cyg X-2, which found that during that source's brightest state, the inner disk radius was poorly constrained yet consistent with being constant \citep{Mondal_etal2018}. In that work, the authors measured the inner disk radius as $12.0^{+3.1}_{-1.6}$ $R_g$ in the fainter, dipping state, and $25.8 \pm10.8$ $R_g$ in the brighter state. As another check on our measurements of the inner disk radius of GX~349+2, we attempted multiple reflection models. While our measurements for the inner disk radius and inclination differed when using \textsc{relxill}, this model did not fit the spectra as well, and yet our main conclusion remains the same --- that the inner disk radius is consistent with remaining unchanged throughout the Z-track. Furthermore, FB3 remains notably different than the other four states, and due to our poor statistical fits with \textsc{relxill} and the aforementioned problems modeling FB3 with \textsc{reflionx$_{BB}$}, we caution against the interpretation that an increase in inner disk radius occurs during intense flaring.

Continuum modeling can provide an alternate and independent measure of the inner disk radius. From the normalization of the disk component in our modeled spectra, we can calculate an inferred value of $R_{in}$. This was done in previous studies of the Fe K$\alpha$ line in GX~349+2, as a method to compare the measured inner disk radius with the value taken from continuum modeling \citep{Cackett2010}. From our reflection modeling, the inferred inner disk radius as calculated by the continuum \textsc{diskbb} normalization decreases from 15.0 km in the NB, to 14.0 km in the vertex, then to 13.1, 11.6, and finally to 9.2 km along the flaring branch, after correcting the apparent radius with a color correction factor of 1.64 --- this factor is a product of the square of the spectral hardening factor $\kappa = 2.0$ with the boundary corrective factor $\xi = 0.41$ \citep{Kubota_etal1998}. A spectral hardening factor of $\kappa = 2.0$ is appropriate assuming the luminosity of the source is close to Eddington \citep{ShimuraTakahara1995}. This clearly does not agree with the inner disk radius as measured from the Fe line, since even after a correction the inferred radii put the disk very near the surface of the neutron star, and also because the clear decreasing pattern described here is not seen in measurements using the Fe line. While this is troubling, it is often the case that the the disk radius inferred from continuum modeling does not agree with the Fe line profile measurements \citep[see][]{Merloni_etal2000, Cackett2010}. Yet another possibility that might explain this discrepancy is that, rather than having a truncated disk, an optically thick corona disrupts the view of the inner accretion disk, which could mean the reflection we detect does not originate in the innermost radii of the accretion disk. An optically thick corona has been suggested for some AGN and LMXBs, though it is typically an optically thin corona that is used to explain Comptonization in LMXBs \citep{Rozanska_etal2015}.

While the inner disk radius is seen to increase dramatically in FB3, this suggestive measurement may not be reliable as confirmation of such a process since the radius measurement is not very well constrained during extreme flaring. Even along the flaring branch, the disk temperature is seen to continue to increase, and one might consider the case when the radiation pressure on the disk surpasses the local Eddington limit and forces the disk out to further radii. However, if the local Eddington limit were reached, we might expect to see a saturation in the disk and blackbody temperatures, rather than what was observed. Furthermore, as has been mentioned before, spectral modeling of FB3 is physically inconsistent, and therefore cannot provide accurate measurements of the physical properties of the system. In hindsight, this failure to fit the spectrum of FB3 is not that startling, considering several characteristics of the flaring in GX~349+2. First, the extreme flaring state FB3 covers a broad range in intensity, across which some change may occur that provides a complicated spectrum after it has been time-averaged. Second, as can be seen in Figure~\ref{LC}, there are 5 distinct flares that are collectively represented by FB3 -- each of which has a different profile and duration, and peaks at a different intensity. As can be seen in Figure~\ref{flares}, the shape of the spectra from each of these flares differs, and they produce notably different best-fit parameter values. The clear distinctions between these individual flares may be the cause of our difficulty in modeling this spectral state. Conversely, our inability to successfully model FB3 might suggest that we do not understand the most extreme flaring of Z sources, or that there may be physical phenomena not represented by the models we have used that must be included to explain such flaring.

Other reflection models were attempted as well, and the best results among these were found using \textsc{relxill}. This model uses a power law spectrum as the incident flux onto the disk to produce reflection. While our measurements for the inner disk radius as well as other parameters depended on the choice of reflection model, our main conclusion remains the same --- that the inner disk radius is consistent with remaining unchanged throughout the Z-track. This appears to be true regardless of the reflection model used.

When all of this is considered, we can only state that the inner disk radius as measured by the Fe K line is roughly unchanged as GX~349+2 moves across its Z-shape, and that a significant change in the geometry of the inner accretion disk may only occur during strong flaring, when such a measurement is not well constrained by our observations. Furthermore, the unusual spectral shape of FB3 suggests that more complicated effects occur during extreme flaring, making it difficult to model with the reflection spectrum that worked so well for all of the source's other states. This motivates further study of GX~349+2 as well as other, similar Z sources during flaring.

\section{Summary}\label{sec:summ}

We present a detailed \textit{NuSTAR} observation of the Sco-like Z source NS LMXB GX~349+2, throughout the Z-track. We split the track into 5 states, which were identified by plotting the hardness versus intensity of the source over the observation, and then defining the states as different regions of the characteristic Z-track. From spectral modeling we detect a significant increase in the temperatures of the blackbody component along the Z-track, from the normal branch through the end of the flaring branch. We also measure the inner disk radius of a truncated accretion disk, which stays more or less consistent in the states of lower flux, and may increase during bright flaring. The brightest flux state, corresponding to the most extreme flaring, proved difficult to model, and our best fits were physically inconsistent. Separating this state into 5 individual flares provided some insight; the distinct flares that collectively represent this brightest flaring state have noticeably different spectral properties, and are best fit by a large range of parameters. Therefore, further study and observation of these bright NS LMXBs, especially during periods of flaring, should be performed. Such observations will be critical in understanding what causes flaring in these sources, as well as the physical properties of these systems very near the neutron star that cause them to trace out their Z-tracks as they move through their various spectral states.

\acknowledgements

We thank Michael Parker for kindly providing the blackbody \textsc{reflionx} model. EMC and BMC gratefully acknowledge support from NASA through the \textit{NuSTAR} Guest Observer Cycle 2, grant NNX17AB66G. RML acknowledges funding through the NASA Earth and Space Science Fellowship.

\bibliographystyle{apj}
\bibliography{apj-jour,benbibliography}

\end{document}